\documentclass{article}

\usepackage{microtype}
\usepackage{graphicx}
\usepackage{subfigure}
\usepackage{booktabs} 

\usepackage{cite}
\usepackage{amsmath,amssymb,amsfonts}
\usepackage{algorithmic}
\usepackage{graphicx}
\usepackage{textcomp}
\usepackage{xcolor}
\def\BibTeX{{\rm B\kern-.05em{\sc i\kern-.025em b}\kern-.08em
    T\kern-.1667em\lower.7ex\hbox{E}\kern-.125emX}}
\usepackage{multirow}
\usepackage{hhline}
\usepackage{amsthm,enumitem}

\usepackage{xspace}

\newcommand{\QuickQuestion}{Quick Question\xspace}

\usepackage{hyperref}

\usepackage[accepted]{icml2020}

\icmltitlerunning{Quick Question: Interrupting Users for Microtasks}

\begin{document}

\twocolumn[
\icmltitle{Quick Question: Interrupting Users for Microtasks with Reinforcement Learning}



\icmlsetsymbol{equal}{*}

\begin{icmlauthorlist}
\icmlauthor{Bo-Jhang Ho}{equal,ucla}
\icmlauthor{Bharathan Balaji}{equal,amazon}
\icmlauthor{Mehmet Koseoglu}{ucla}
\icmlauthor{Sandeep Sandha}{ucla}
\icmlauthor{Siyou Pei}{ucla}
\icmlauthor{Mani Srivastava}{ucla}
\end{icmlauthorlist}

\icmlaffiliation{ucla}{University of California, Los Angeles}
\icmlaffiliation{amazon}{Amazon, Seattle}

\icmlcorrespondingauthor{Bo-Jhang Ho}{bojhang@ucla.edu}

\icmlkeywords{Machine Learning, ICML}

\vskip 0.3in
]



\printAffiliationsAndNotice{\icmlEqualContribution} 

\begin{abstract}
Human attention is a scarce resource in modern computing. A multitude of microtasks vie for user attention to crowdsource information, perform momentary assessments, personalize services, and execute actions with a single touch. A lot gets done when these tasks take up the invisible free moments of the day. However, an interruption at an inappropriate time degrades productivity and causes annoyance. Prior works have exploited contextual cues and behavioral data to identify interruptibility for microtasks with much success. With Quick Question, we explore use of reinforcement learning (RL) to schedule microtasks while minimizing user annoyance and compare its performance with supervised learning. We model the problem as a Markov decision process and use Advantage Actor Critic algorithm to identify interruptible moments based on context and history of user interactions. In our 5-week, 30-participant study, we compare the proposed RL algorithm against supervised learning methods. While the mean number of responses between both methods is commensurate, RL is more effective at avoiding dismissal of notifications and improves user experience over time.
\end{abstract}

\section{Introduction}
Human computer interaction has evolved over the years from desktop-only machines to wearables that interface at a glance. Modern services in navigation, local business discovery, crowdsourcing, 
participatory medicine~\cite{lejbkowicz2010internet} depend upon such on-demand interaction, where the user can access the services wherever they go. Push notifications exploit this interaction to proactively seek user attention. Notifications are used to check mail, remind users, nudge behavior, 
get feedback\footnote{Yelp review - https://www.yelp.com/}, label datasets\footnote{Google Maps: Question About a Place - https://goo.gl/Jf9mTq}, etc. However, human attention is a limited resource~\cite{lee2015case}, and serving content irrelevant to the context leads to annoyance, reduces productivity~\cite{bailey2006need} and diminishes engagement~\cite{mehrotra2016my}.



User interruptibility has been extensively studied in literature~\cite{mehrotra2017intelligent}. 
We categorize prior works into two approaches - rule-based and data-based policies. The rule-based policy relies on human behavior analysis and identifies the moments that people are likely available. Proposed policies include identifying breakpoints between two tasks~\cite{iqbal2010oasis} and using events such as unlocking the phone as a heuristic~\cite{vaish2014twitch}. 
As the policy is fixed, it does not fit users who have different preferences. Data-based approach leverages machine learning~\cite{pejovic2014interruptme, mehrotra2015designing, sarker2014assessing, pielot2015attention}. Prior works used supervised learning that learns the non-linear relationships between user context and availability using dataset collected from an existing policy. 
Thus, the data-based policy learns preferences for each user based on behavior. 

While prior works use supervised learning (SL), we propose using reinforcement learning (RL) to identify user interruptibility. We identify the following advantages of RL: \\
\textbf{(i) Sequential decision process:} SL assumes data samples are independent from each other, whereas RL models each sample a function of previous samples. RL can learn that users will get annoyed if they get too many notifications. \\
\textbf{(ii) Exploration:} SL methods passively collect data based on an existing policy, while RL algorithms actively explore the problem space to learn policies that are robust.\\
\textbf{(iii) Online learning:} SL methods need a training dataset to learn whereas RL is designed for online learning. 



We focus on identifying interruptibility for microtasks~\cite{cheng2015break}, where we ask the user a ``quick question'' that can be answered in a few seconds. Microtasks have several use cases - crowdsourcing, personalization~\cite{organisciak2014crowd}, labeling datasets~\cite{good2014microtask}, ecological momentary assessment~\cite{ponnada2017microinteraction}. We seek to identify appropriate moments of the day to maximize microtask responses. We collect user context using a smartphone and periodically send the information to the cloud. Our web server uses SL and RL to determine whether to send a microtask to user. User interactions over time are used to train the models.

We conducted a 5-week, 30-participant user study to compare SL and RL methods. Our results indicate the microtask responses vary dramatically from person to person and both data-based methods capture the individual preferences. We penalized notification dismissals with a negative reward for RL, and it effectively learned to avoid dismissals. However, the number of responses is higher for SL. Users indicated they were available to answer quick questions when the RL agent interrupted them 73\% of the time compared to 54\% for SL. Users expressed improved experience over time with RL and data indicates that RL adapts to changing preferences within a few days.


The following are the contributions of this work:
\begin{itemize}
\itemsep0em 
  \item We implemented a cloud service that collects user context from a smartphone app and determines interruptibility using both supervised learning and reinforcement learning.
  \item We conducted a 5-week user study and recruited 30 participants to compare supervised-learning and reinforcement-learning based microtask scheduling. 
\end{itemize}

\section{Related Work}

\subsection{Microtask}

A \emph{microtask} typically refers to a simple task that can be done within seconds~\cite{cheng2015break}. The microtask technique is widely used in crowdsourcing context: This technique aims to lower the mental burden~\cite{kittur2008crowdsourcing} and to improve response quality~\cite{cheng2015break}. Microtasks have also been applied to solve big, complex tasks by partitioning them into multiple independent microtasks~\cite{kittur2011crowdforge}. A more sophisticated approach is to automatically decompose a task based on domain ontology~\cite{luz2014generating}. Microtask techniques have been successfully applied to high-complexity tasks such as 
article writing~\cite{kittur2011crowdforge} and software development~\cite{latoza2014microtask}. In this paper, we address an orthogonal issue how to schedule microtasks to increase user responses.

\subsection{Interruptibility Modeling}

Machine-to-human interruptibility has been studied extensively. 
One major interruption source from mobile and wearable devices is push notifications; prior studies have shown that scheduling notifications at an improper time increases anxiety~\cite{pielot2014situ} and reduces productivity~\cite{bailey2006need}. We broadly categorize interruptibility modeling techniques into rule-based and data-based. Rule-based techniques rely on prior knowledge to estimate opportune moments of interacting with people. For example, opportune moments can be identified based on mobile phone usage pattern such as after phones are unlocked~\cite{vaish2014twitch}, after phone calls or text messages are finished~\cite{fischer2011investigating}, or when a user \emph{reviews} an application~\cite{banovic2014proactivetasks}. Scheduling microtasks at the task boundaries (i.e. \emph{breakpoints}) can reduce mental effort~\cite{adamczyk2004if}, and this technique has been applied to the desktop domain~\cite{iqbal2010oasis} and mobile platforms~\cite{okoshi2015reducing}. Goyal et al.~\cite{goyal2017intelligent} observes that it would be the most effective to schedule a notification when electrodermal activity increases under a high-stress context. All these techniques rely on a fixed policy and cannot be generalized to all users or adapted for changes in user preference.

The data based approaches derive a classification model based on the user context, which is sensed by mobile or wearable devices. Besides the common mobile phone sensor data such as time, location, and motion activity, ~\citet{sarker2014assessing} further consider stress level and social engagement, and uses SVM to detect when a user is available. InterruptMe~\cite{pejovic2014interruptme} takes emotion as an additional feature to infer if sending an instant message is appropriate at the moment. ~\citet{mehrotra2015designing} leverage the content of notifications to infer how likely the notifications will be responded. ~\citet{pielot2015attention} deliver news feeds when a user gets bored by training a random forest classifier. PrefMiner~\cite{mehrotra2016prefminer} mines the notification usage patterns and users can pick some of those patterns to effectively filter out undesired notifications.
Thyme~\cite{aminikhanghahi2017thyme} shares the same goal with us to maximize user responses to microtasks. They use SVM to identify interruptibility. Our work differs from these works by applying reinforcement learning techniques to address the interruptibility problem. As opposed to supervised learning, reinforcement learning is an online learning process and it learns user preference from interacting with users without a separate training phase.

\subsection{Human-in-the-Loop Reinforcement Learning}

Reinforcement learning (RL) has recently achieved state of the art performance in domains such as games~\cite{mnih2015human,silver2017mastering} and robotics~\cite{gu2017deep,andrychowicz2020learning}. 
These breakthroughs demonstrate the capability of reinforcement learning. There are several works that apply RL to help humans. Sentio~\cite{salma2018Sentio} uses a variant of Q-learning to prompt forward collision warnings in cars. ~\citet{rafferty2016faster} develop a tutor system based on Partially Observable Markov Decision Process. ~\citet{greenewald2017action} exploit contextual bandit to enhance a mobile health system. ~\citet{silver2013concurrent} use RL to maximize an objective of a company (e.g., revenue) by performing actions to customers (e.g., offering a discount). Our work aligns with these works and uses RL to optimize notification response performance.

\section{Methods}

\begin{figure}[!t]
  \begin{center}
    \centerline{\includegraphics[width=0.38\textwidth]{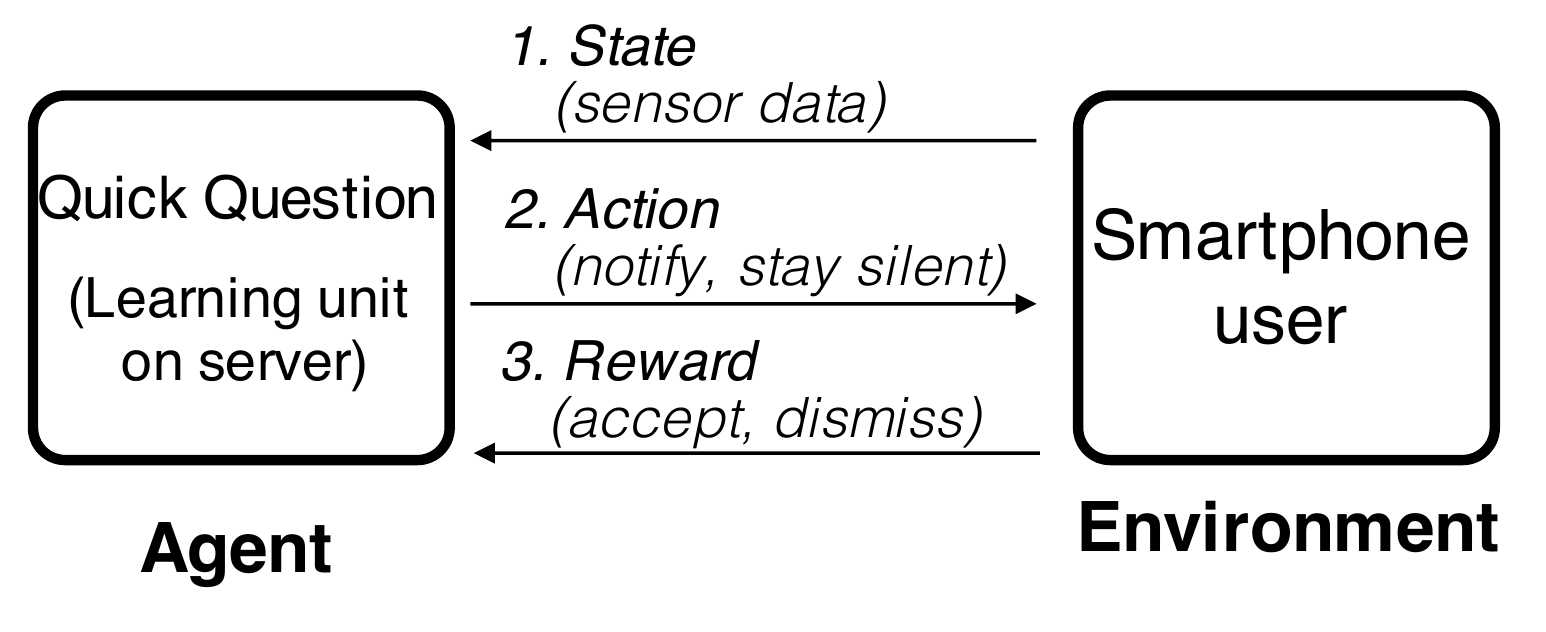}}
    \vspace{-0.5cm}
    \caption{Reinforcement learning setup.}
    \label{fig:reinforcement_learning_explanation}
  \end{center}
  \vspace{-1em}
\end{figure}

In our partially observed Markov Decision Process setting, there is an \emph{agent} and an \emph{environment} whose relationship is depicted in Figure~\ref{fig:reinforcement_learning_explanation}. At each step, the agent first makes an \emph{observation} to obtain a representation of the environment called \emph{state}. The observation is an approximate representation of state. The agent then takes an \emph{action} based on its \emph{policy}. As a result of the action, the environment moves to a new state and returns a \emph{reward}. The agent maximizes the discounted sum of future rewards accumulated over successive steps. \QuickQuestion uses the Advantage Actor-Critic (A2C)~\cite{mnih2016asynchronous} as the RL algorithm. We describe the algorithm details in Appendix \ref{appendix:algorithm}.


In our framework, the \emph{agent} is our \textbf{\QuickQuestion} system, and the \emph{environment} is the smartphone user. The agent observes the \textbf{user context} as a representation of user \emph{state}, and takes an \emph{action}:  either \textbf{to send a notification}, or \textbf{to keep silent}. The agent gets a positive reward when the notification is answered, and a negative reward when the notification is dismissed. A separate agent is trained for each user.

\begin{table}[!tb]
  \caption{Features considered in \QuickQuestion as user context.}
  \scriptsize
  \begin{center}
    \begin{tabular}{|c|c|l|}
      \hline
      \textbf{Sensing modality} & \textbf{Category} & \multicolumn{1}{|c|}{\textbf{Values}}
 \\
      \hline
      Time of the day & Continuous & 00:00 to 23:59 \\
      \hline
      Day of the week & Continuous & Sunday (0) to Saturday (6) \\
      \hline
      Location & Discrete & {Home, Work, Others} \\
      \hline
      \multirow{2}{*}{Motion}  & \multirow{2}{*}{Discrete} & Stationary, Walking, Running, \\
        &  & Biking, Driving \\
      \hline
      Ringtone mode & Discrete & Silent, Vibration, Normal \\
      \hline
      Screen mode & Discrete & On, Off \\
      \hline
      Notification elapsed time$^*$ & Continuous & 0 to 120 (minutes) \\
      \hline
      \multicolumn{3}{l}{$^{\mathrm{*}}$Defined as how many minutes has elapsed since last notification.}
    \end{tabular}
    \label{tab1e:feature_list}
  \end{center}
  \vspace{-1em}
\end{table}

\QuickQuestion consists of a phone app and a web server. 
The phone app senses user context data, and sends it to the server every minute. The server determines if the user is interruptible at the moment based on user context, and sends the decision to the phone app. The phone app collects the following contextual features: time of the day, day of the week, location, motion status, screen status, ringer mode and the elapsed time since last notification (Table \ref{tab1e:feature_list}).

\begin{table*}[!htbp]
  \caption{\QuickQuestion client app includes 9 types of microtasks. The examples of each microtask are provided below.}
  \scriptsize
  \begin{center}
    \begin{tabular}{|c|c|l|c|c|}
      \hline
        &   &   & \textbf{Number of} & \textbf{With gold} \\
      \textbf{Category} & \textbf{Microtask} & \textbf{Example question statement \textit{(options)}} & \textbf{questions} & \textbf{standard} \\
      \hline
      Self- & Availability & Are you available at the moment? \textit{(Yes, No)} & 1 & No \\
      \cline{2-5} 
      monitoring & Emotion & Which describe your current emotion? \textit{(Stressed, Neutral, Relaxed)} & 4 & No \\
      \cline{2-5} 
      & Hydro Diary & How long ago did you drink water? \textit{(Within 1 hour, Within 2 hours, longer)} & 1 & No \\
      \cline{2-5} 
      & \ \ Diet Tracker $^a$ & Do you regularly eat wholegrain cereals, with no added sugar? \textit{(Yes, No)} & 30 & No \\
      \cline{2-5} 
      & Planning & Where are you going after you leave here? \textit{(Home, Work, Others)} & 3 & (Yes)$^c$ \\
      \hline
      Participatory & Noise level & How loud is it at your location? \textit{(Loud, Moderate, Quiet)} & 1 & No \\
      \cline{2-5} 
      sensing & Crowdedness & How many people are there around you? \textit{(0-5, 6-20, $>$20)} & 1 & No \\
      \hline
      Crowd- & \ \  Image labeling $^b$ & What is the object in the image? \textit{(Bear, Bird, Butterfly)} & 1,100 & Yes \\
      \cline{2-5} 
      sourcing & Arithmetic & What is answer of $23 \times 33$? \textit{(259, 759, 1259)} & 1,000 & Yes \\
      \hline
      \multicolumn{5}{l}{$^a$ Source: How healthy is your diet Questionnaire. https://tinyurl.com/y9hssxz7} \\
      \multicolumn{5}{l}{$^b$ Source: ImageNet dataset~\cite{deng2009imagenet}.} \\
      \multicolumn{5}{l}{$^c$ This question is not considered for analyzing user response accuracy due to potential sensor error.}
    \end{tabular}
    \label{tab1e:question_list}
  \end{center}
  \vspace{-1em}
\end{table*}

We use ecological momentary assessment (EMA) questions, commonly used in human behavioral studies~\cite{stone1994ecological}. All the questions are designed in the $\mu$-EMA style~\cite{ponnada2017microinteraction} in multiple choice form and can be answered within a few seconds. The questions can be partitioned into: (i) Self-monitoring questions that track user's mental and physical status such as stress and diet; (ii) Participatory sensing questions collect the environmental information, e.g., noise level at the current location; (iii) The crowdsourcing questions, e.g., image ground truth labeling. We have nine question types listed in Table~\ref{tab1e:question_list}. Some questions are factual, so we can verify user responses.

\begin{figure}[t!]
  \centerline{\includegraphics[width=0.33\textwidth]{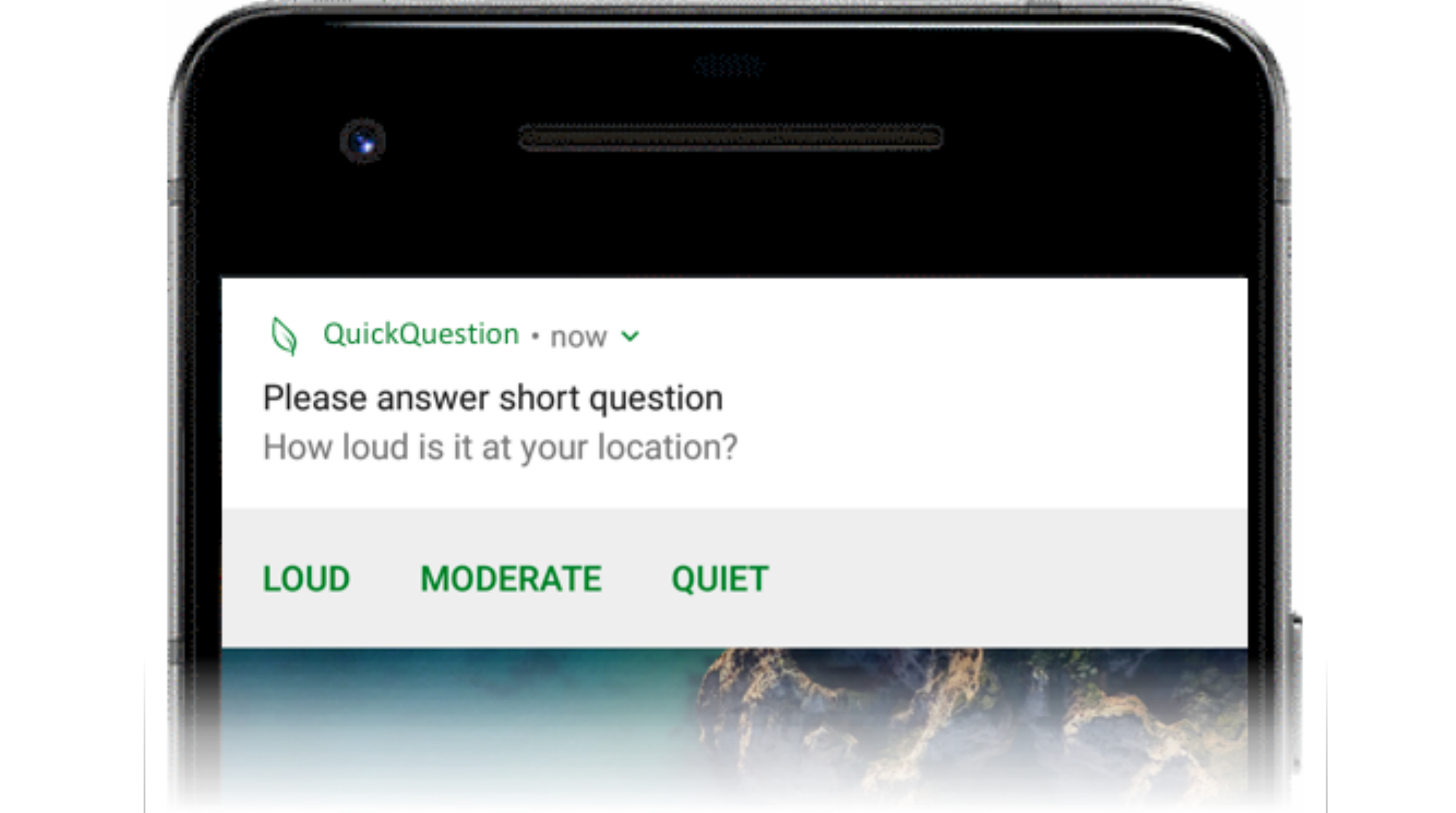}}
  \caption{The microtask answering interface in the client app, through a notification (left) or in the app (right).}
  \label{fig:screenshot}
  \vspace{-1em}
\end{figure}

We embed the questions into the push notification (Figure~\ref{fig:screenshot}). The notification is displayed \emph{heads-up} style with the possible options right below. Alternatively, users can also answer the questions by manually launching our app and selecting a choice in the task view. A microtask times out after an hour, or when a new microtask is scheduled.

\paragraph{Supervised Learning (SL) Agent} converts the user context into a feature vector and the user response as a classification label. We normalize the sensing modalities which output a continuous value (e.g., time of the day) into a number between 0 and 1, and use one-hot encoding to represent sensors with discrete values. We create a positive label if the notification is answered, and a negative label otherwise.

There are two stages in SL: A training phase for data collection and a testing phase. In the training phase, the agent randomly decides whether to send a notification every $\tau$ time units (e.g. every 30 minutes). The training phase lasts for three weeks\footnote{Prior work on training personal models using SL demonstrates that the classification accuracy converges in two weeks, e.g.,~\cite{pejovic2014interruptme, mehrotra2015designing}}. The agent trains a classifier before moving into the testing phase. We use Random Forest as our supervised learning algorithm because it outperforms Support Vector Machine and Neural Networks in our empirical study. Our implementation uses the Scikit-Learn library~\cite{scikit-learn}. 


\paragraph{Reinforcement Learning (RL) Agent} uses the same feature representation as SL to encode user context. RL maps the user response to different reward values: A positive reward that decays exponentially based on response time to encourage scheduling a microtask that can get an immediate response; a strong negative reward if a user dismisses the notification to avoid negative user experience; a small penalty if the user ignores the notification (i.e., does not answer it within one hour because the user overlooks it or forgets to reply). We define the reward function as: 
\[
    reward = 
\begin{cases}
    1 \times t^{0.9}, & \text{if answered,}\\
    -0.1,    & \text{if ignored,}\\
    -5,      & \text{if dismissed}
\end{cases}
\]

where $t$ is the notification response time (i.e., the time difference between the prompt and when the answer is received). 

Our RL agent implementation is built upon Coach~\cite{caspi_itai_2017_1134899}, a reinforcement learning library. Our RL algorithm is selected based on an empirical study with simulated users. We tested neural-network based RL algorithms including Deep Q-learning Network (DQN)~\cite{mnih2015human}, Advantage Actor-Critic (A2C)~\cite{mnih2016asynchronous}, and Proximal Policy Optimization (PPO)~\cite{schulman2017proximal}. A2C achieves the best performance among these algorithms and converges in the shortest time. Hence, we choose A2C for the real user study. We employ a fully-connected neural network with one hidden layer (256 units). We set the discount factor $\gamma=0.99$. The algorithm uses categorical exploration strategy which performs a stochastic action based on the probability distribution of actions.
We include system implementation details in Appendix \ref{appendix:system}. We list the hyperparameters used in Appendix \ref{appendix:hyperparameters}.


\section{User Study}
We were guided by the following inquiries in our user study:
\begin{itemize}
\itemsep0em
  \item What was the relative \emph{notification response amount, rate, and accuracy} (for notifications with correct answers) collected from the reinforcement learning (RL) method and how is it compared to the supervised learning (SL) method?
  \item How did the \emph{user experience} resulting from the RL method compare to the SL method?
\end{itemize}

To make a meaningful comparison between SL and RL, we maintain consistency in confounding factors, such as the user study procedure, the qualification criteria for participants, and the analysis method on the collected notifications. 


\subsection{Participants}
This study was approved by UCLA IRB (IRB\#18-000504). In total, we recruited 30 participants (19 females, 11 males) from a major research university. Among these participants, 28 were students and 2 were staff. Ages ranged from 17 to 29 (mean=21.1). The inclusion criteria of our study are active Android users with OS version 7 or higher. 
Participant phone models included Samsung (N=11), Google Pixel/Nexus (N=8), OnePlus (N=6), Sony (N=2), LG (N=2), and HTC (N=1). Additionally, Android OS 7, 8, 9 accounted for 7, 18, and 5 participants respectively. The participants received gratuity of \$50 for each completed week, and an additional \$50 if they complete the entire study. 
  

\subsection{Procedure}
15 participants were part of the RL group, and 15 were in the SL group. 
The procedure consists of two phases: (1) a screening phase to select qualified participants, and (2) an experimental phase. Participants were recruited via university mailing lists and snowball sampling. 

In the screening phase, interested candidates completed a questionnaire regarding the phone model they were using, its OS version, and whether they would have network reception during the entire study even if WiFi is not available. After they passed the screening phase, candidates were asked to fill out a pre-study questionnaire with their personal information. Finally, qualified participants were asked to attend an orientation on how to use the \QuickQuestion app. During the orientation, we emphasized that (1) our study app will send no more than 150 notifications\footnote{To test the limit of how many notifications one can handle, we choose a number twice larger than the number of daily notifications (53.8)~\cite{pielot2018dismissed}.} in each day between 10am to 10pm, and each question can be answered within a few seconds, and (2) participants were asked to not change the way they respond to notifications, hence, answering all questions is not necessary. We then helped the participants to install our app and complete the location configuration (i.e., user's home and work location) for the classifier. 

In the 5 week experimental phase, participants went about their everyday activities with their app-installed phone. We sent a weekly survey at the end of each week to gauge user perception towards the notification schedule on a 1-5 point Likert scale. At the end of 5 weeks, we conducted a post-study survey consisting of open-ended questions to gather participant feedback on the overall user experience.

\section{Evaluation}

\begin{table*}[!htbp]
  \scriptsize
  \caption{The comparison of the response performance of short questions between RL and SL algorithms. There are 15 participants in each group}
  \begin{center}
    \begin{tabular}{ c  cccc || c  cccc }
      \hline
      \multicolumn{5}{c||}{\textbf{Reinforcement learning algorithm}} & \multicolumn{5}{c}{ \textbf{Supervised learning algorithm}} \\
      \hline
       & \# answered   & \# dismissed & Answer & Dismiss  &  &  \# answered   & \# dismissed & Answer & Dismiss  \\
       & notifications & notification & rate   & rate  &  & notifications & notification & rate   & rate  \\
      \hline
\textbf{Week 1} & 214.9 & 11.8 & 0.42 & 0.03 &  \textbf{Week 1} & 88.6 & 5.5 & 0.55 & 0.04 \\
      \textbf{Week 2} & 182.2 & 25.4 & 0.28 & 0.04 &  \textbf{Week 2} & 108.3 & 12.0 & 0.45 & 0.06 \\
      \textbf{Week 3} & 138.6 & 22.3 & 0.22 & 0.06 &  \textbf{Week 3} & 128.6 & 13.1 & 0.43 & 0.05 \\
      \cline{6-10}
      &  &  &  &  & \textbf{Avg (SL-train)} & 108.5 & 10.2 & 0.48 & 0.05 \\
      \hhline{~~~~~||=====}
      \textbf{Week 4} & 120.1 & 9.0 & 0.21 & 0.02 &  \textbf{Week 4} & 155.7 & 19.2 & 0.36 & 0.04  \\
      \textbf{Week 5} & 112.2 & 8.1 & 0.20 & 0.02 &  \textbf{Week 5} & 173.4 & 24.6 & 0.32 & 0.07  \\
      \hline
      \textbf{Avg (RL)} & 153.6 & 15.3 & 0.27 & 0.03 &  \textbf{Avg (SL-test)} & 164.6 & 21.9 & 0.34 & 0.06 \\
      
      \hline
    \end{tabular}
    \label{tab1e:response_performance}
  \end{center}
  \vspace{-1em}
\end{table*}

In total, we collected 19,223 hours of data. 
Among these data, our system sent out 66,300 notifications. 20,950 (31.5\%) notifications were answered, 2,008 (3.0\%) were dismissed, and 43,342 (65.5\%) were ignored. We compare the performance of RL and SL from different dimensions. In SL, we report the results in the training phase (SL-train) and the testing phase (SL-test) separately.

%
%
%

\subsection{Task Response}

Table~\ref{tab1e:response_performance} compares the task response performance. We list four metrics in the table: the number of notifications answered in a week, the number of notifications dismissed in a week, the ratio of answered notifications, and the ratio of dismissed notifications.

\textbf{RL gets more microtasks answered, but SL achieves better answer rate.} On average, RL is able to get 153.6 microtasks answered per week, which is slightly higher than SL with 130.9 microtasks per week. However when we break it down, RL only outperforms SL-train (108.5/week) but is slightly lower than SL-test (164.6/week). SL achieves a higher answer rate (48\% in SL-train and 34\% in SL-test) than RL (27\%). Therefore, SL is effective in learning interruptibility as indicated by prior studies. However, the random notification schedule in the training phase underestimates user responsiveness and re-training the SL model more frequently could have led to improved results.

The reward function of RL incentivizes answering microtasks and discourages dismissals with a heavy penalty. However, the penalty for an ignored microtask is low. In addition, we only show one micro-task at a time, the prior micro-task notifications are removed before sending a new one. Given this design, RL agent learned to send lots of microtasks because users ignored most notifications but dismissed very few of them. Hence, RL agent gets a lot of questions answered, but its response rate is low.

\begin{figure}[t!]
  \centerline{\includegraphics[width=0.49\textwidth]{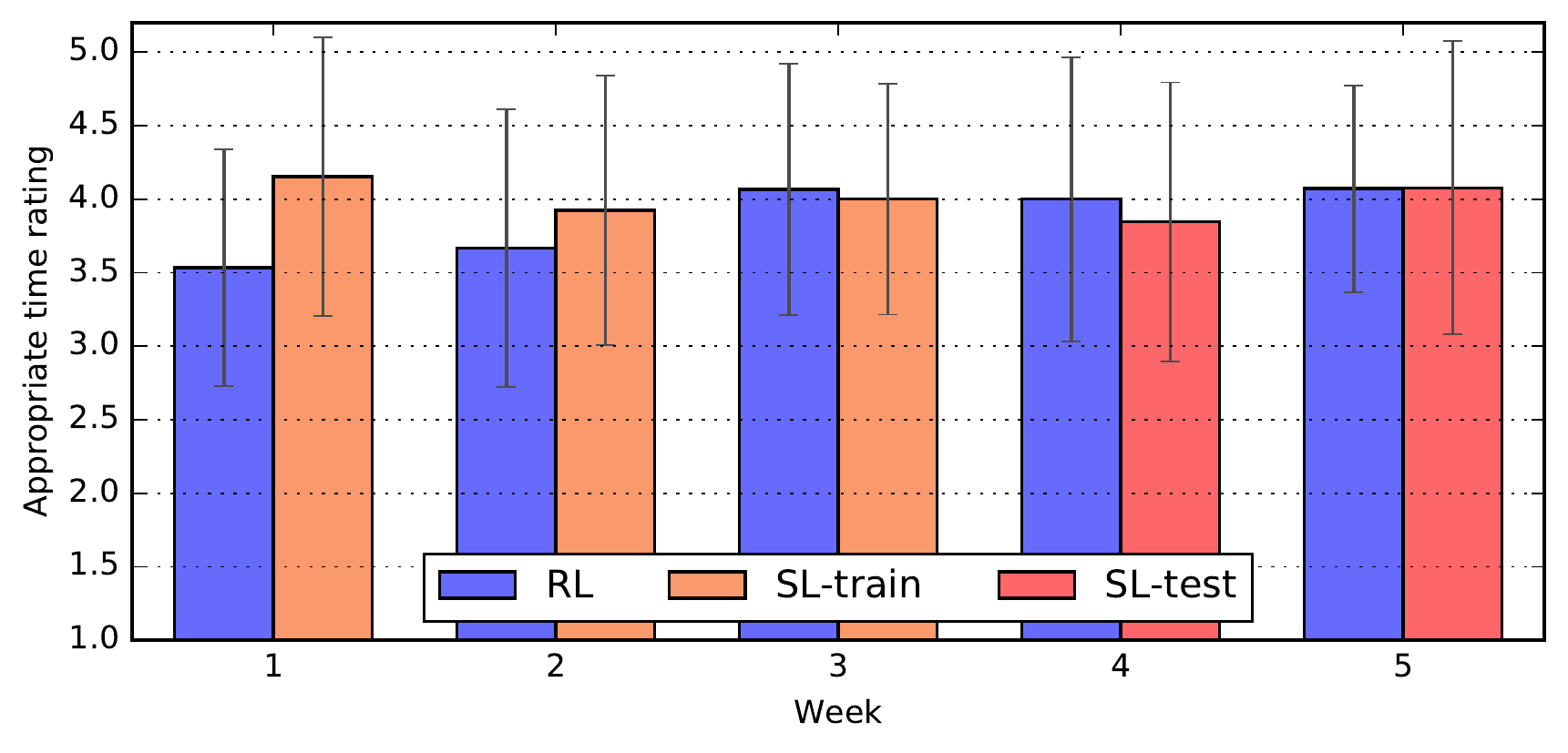}}
  \vspace{-0.5cm}
  \caption{Weekly rating of both algorithms.}
  \label{fig:weekly_survey_rating}
  \vspace{-2em}
\end{figure}

\textbf{RL can effectively suppress dismissed notification rate.} RL keeps the task dismiss rate low (3\%). In contrast, both SL-train and SL-test exhibit higher dismiss rate (6\%). Note that RL agent's dismiss rate is low despite the fact that it sends larger number of notifications. RL has been incentivized to avoid dismisses with a sequential decision making process, whereas the loss function in vanilla SL algorithms pick actions independently based on probability distribution of past data and do not learn the impact of their actions on the user state. This makes decisions made by SL-test sticky, and increases the notification volume of SL-test 2.1x more than SL-train. One can potentially improve SL performance using recurrent neural models.

\textbf{Rewards have high variance across users.} RL receives higher weekly rewards ($43.2\pm136.6$) than SL ($26.4\pm274.2$), but reward pattern vary widely across users. Hence, we make no statistically significant conclusion w.r.t rewards.

\begin{figure*}[!t]
  \begin{minipage}[t]{.49\textwidth}
    \centering
    \includegraphics[width=1\textwidth]{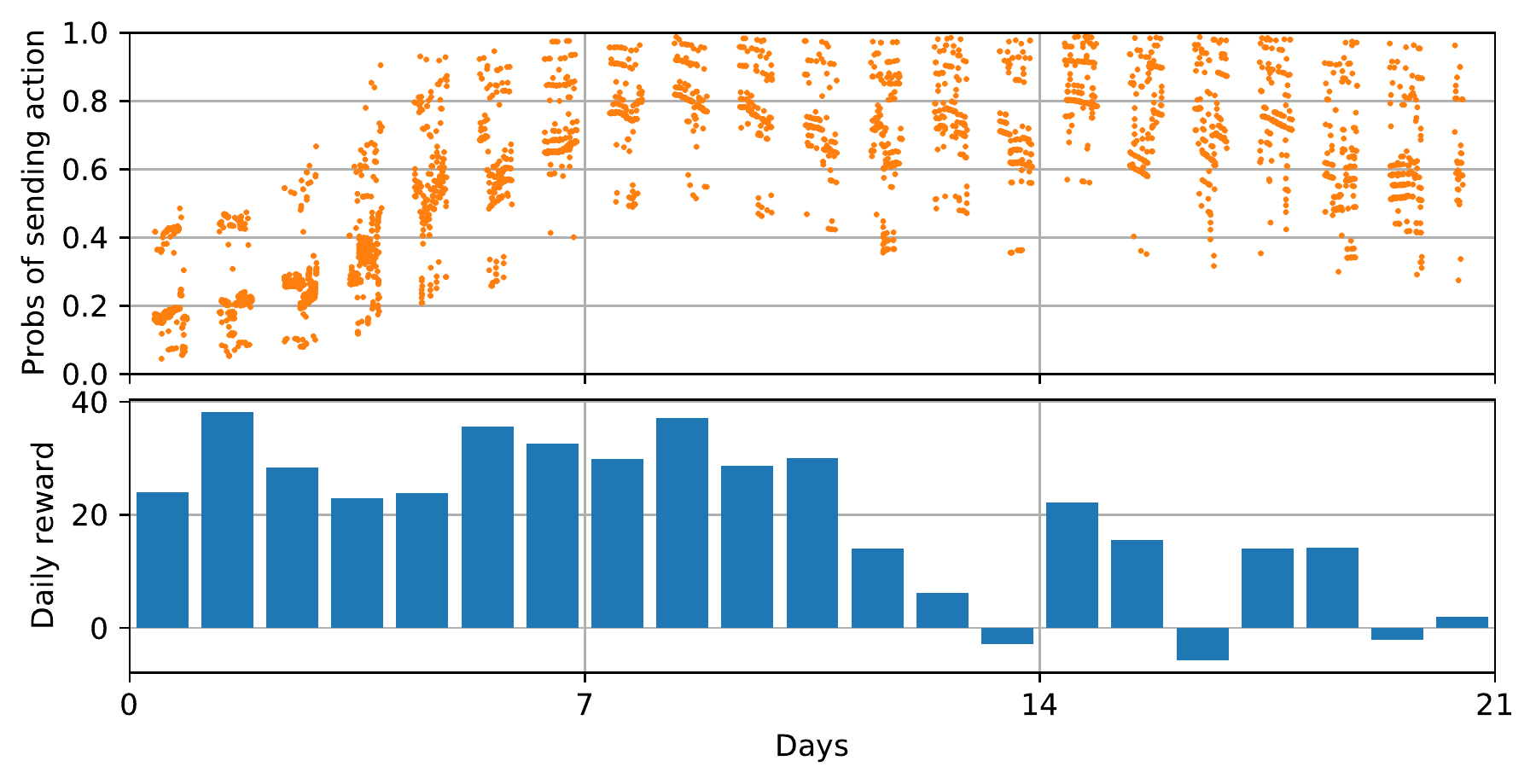}
    \small (a) The RL agent takes one week to converge.
  \end{minipage}
  \begin{minipage}[t]{.49\textwidth}
    \centering
    \includegraphics[width=1\textwidth]{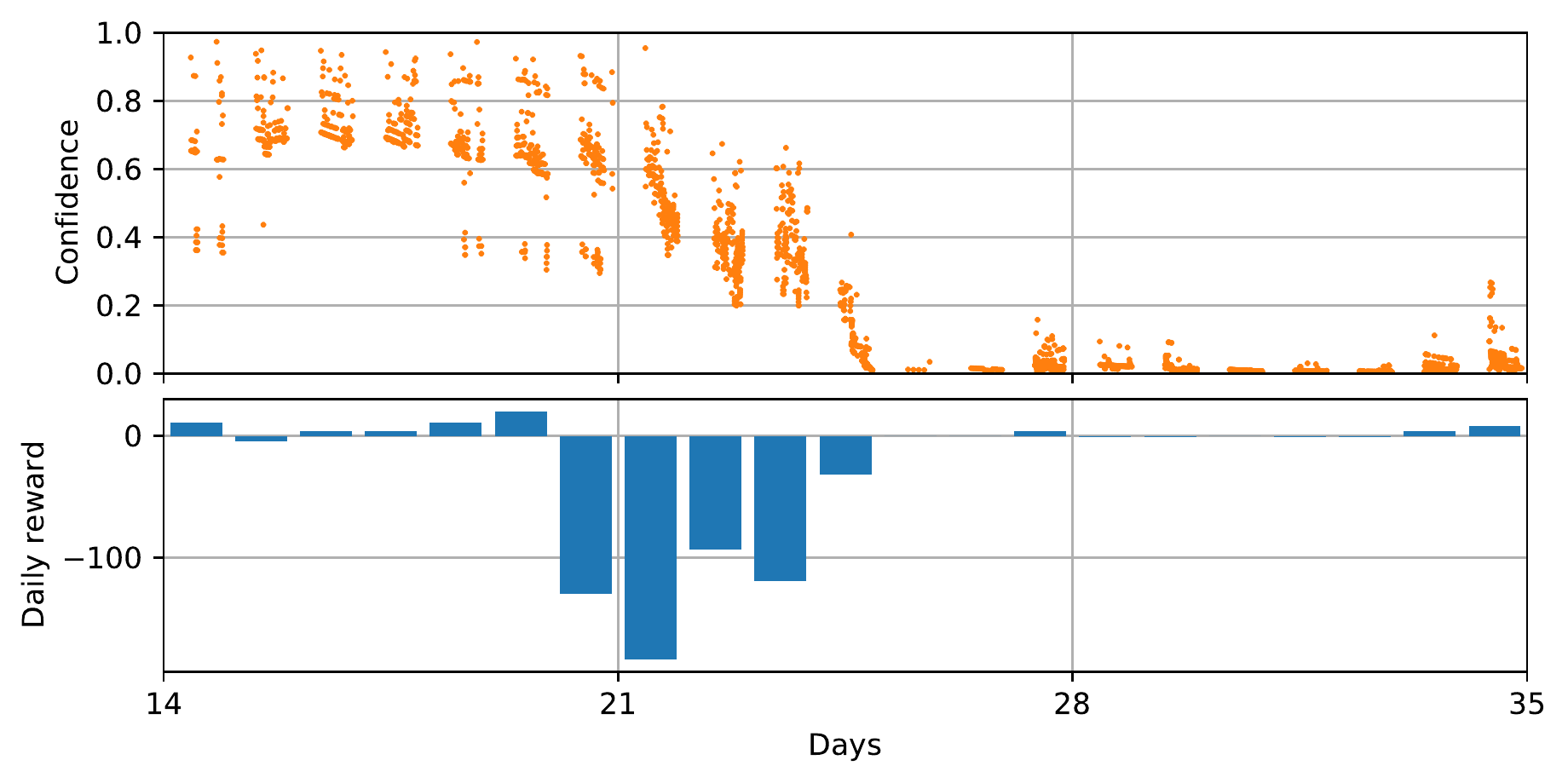}
    \small (b) The RL agent can quickly adapt to user preference change.
  \end{minipage}
  
  \caption{Probability of sending a microtask across time by the reinforcement learning (RL) agent for two different users.}
  \label{fig:action_probability_individual}
  \vspace{-1em}
\end{figure*}

\subsection{User Experience of Interruptibility}

\textbf{User experience in RL is initially worse, but improves over time.} Figure~\ref{fig:weekly_survey_rating} shows the weekly survey result in which we ask participants to rate the appropriateness of the timing of the prompted tasks with a 5 point Likert scale. The result shows that SL starts with a high rating ($4.0 \pm 0.9$ in SL-train), and the rating remains relatively flat in the testing phase ($4.0 \pm 1.0$ in SL-test). RL starts with a low rating ($3.6 \pm 0.9$ in the first two weeks). This is likely due to the fact that RL sends more notifications to explore the problem space, and this causes disturbance. The rating in RL improves over weeks ($4.0 \pm 0.8$ in the $5^{th}$ week). 

\subsection{Microtask Response Analysis}


\begin{figure}[!tbp]
  \begin{minipage}[b]{.23\textwidth}
    \centering
    \includegraphics[width=1\textwidth]{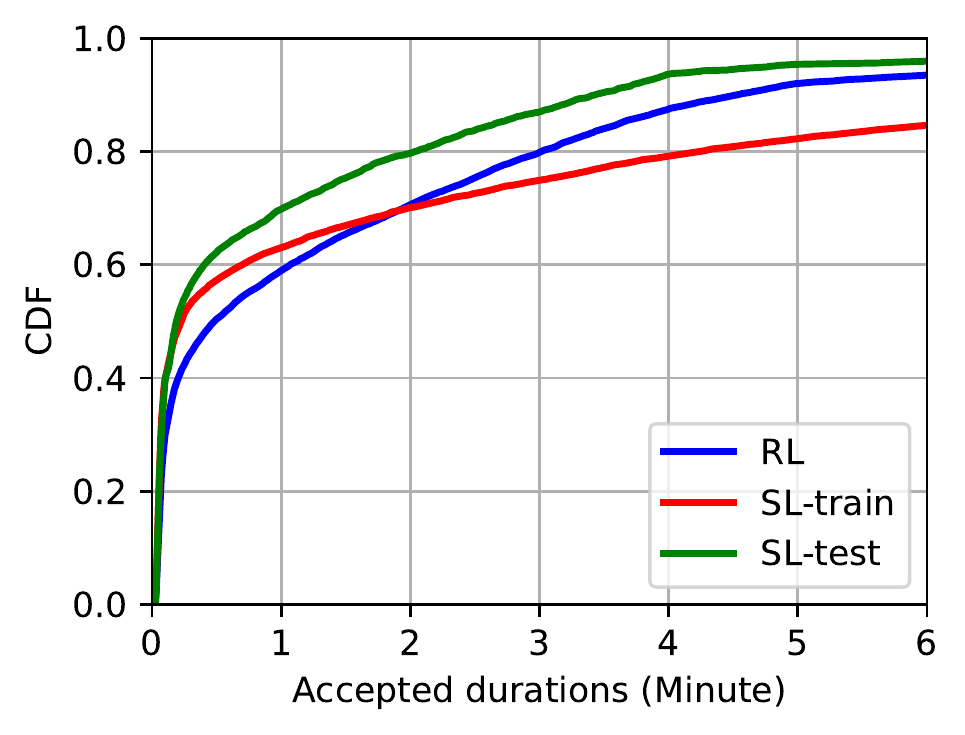}
    (a)
  \end{minipage}
  \begin{minipage}[b]{.23\textwidth}
    \centering
    \includegraphics[width=1\textwidth]{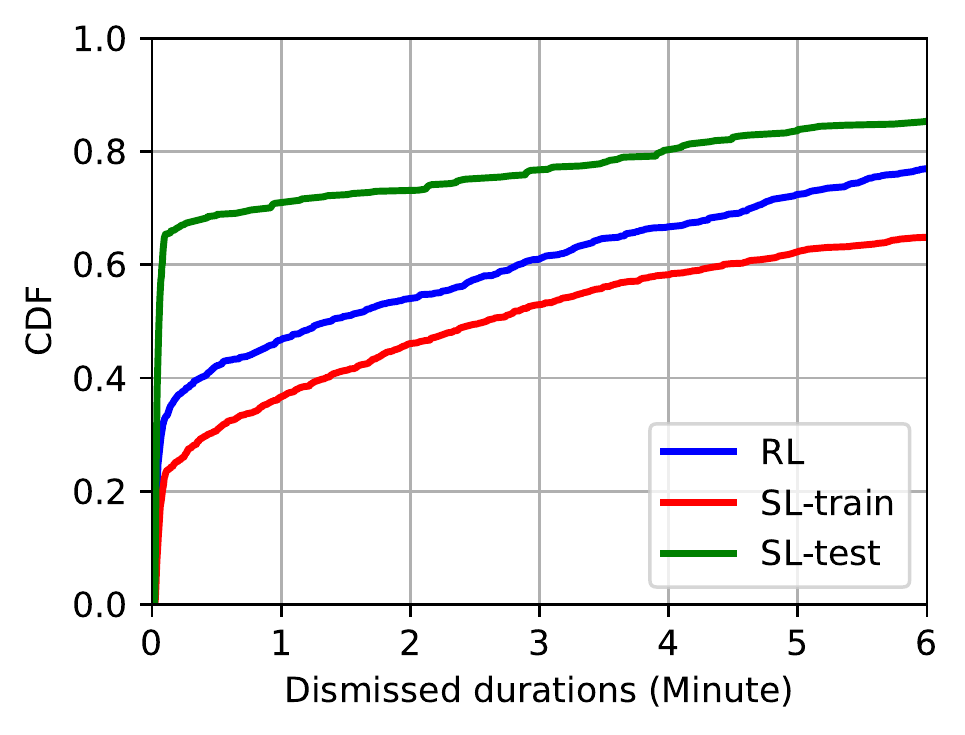}
    (b)
  \end{minipage}

  \caption{CDF of time intervals for (a) answering and (b) dismissing notifications.}
  \label{fig:notification_reaction_durations}
  \vspace{-1em}
\end{figure}

\textbf{RL can better identify available moments.} 
~\citet{pielot2017beyond} identified that users respond to microtasks even when their perception is that they are not available. Our result shows that 73\% of the responses from RL participants are \emph{yes} which indicate the users were available when they answered the questions. However, only 54\% of the responses in SL indicates users were available, suggesting that RL does a better job of finding interruptible moments.  


\textbf{People dismiss notifications much faster in SL-test.} Another measure of interruptibility can be to observe how long users take to respond to microtasks. Figure~\ref{fig:notification_reaction_durations}a displays the time intervals that users take to answer a notification since prompted. The result shows that at least 58\% of the microtasks are answered within one minute in both algorithms. 
Figure~\ref{fig:notification_reaction_durations}b shows that 34\%, 24\%, and 65\% of the notifications are instantly dismissed in RL, SL-train, and SL-test (i.e., within 5 seconds) right after they are scheduled. 

\textbf{High response accuracy.} 
We define \emph{response accuracy} as number of correctly answered microtasks over the number of factual tasks (Table~\ref{tab1e:question_list}). Both algorithms achieve over 90\% of response accuracy in all the 5 weeks, validating that the participants were engaged throughout the study.

\subsection{Learning Algorithm Analysis}




\textbf{A2C converges in a week.} To understand when the RL agent starts to learn something meaningful, we pick one user as an example and plot the confidence scores of all the interruptibility queries in Figure~\ref{fig:action_probability_individual}a. The \emph{confidence} is defined as the likelihood for the learning algorithm to send a microtask, which is part of the output of A2C. We provide daily reward on the bottom for comparison. Since the agent receives bigger rewards for the first 7 days
, the agent gets more confident in prompting notifications. As the user behavior changes around $10^{th}$ day, the daily reward and the confidence drops. It can be explained that as time progresses, the agent adapts to the changing user behavior. 


\textbf{RL can adapt to user preference change and capture the weekly pattern.} Figure~\ref{fig:action_probability_individual}b presents another user who actively dismissed notifications in the middle of the study. The amount of dismissed notifications significantly increased after day 20. The confidence drops when RL starts receiving negative reward which in turn suppresses the microtasks to this user. However, the confidence raises on day 28 and day 35 which are Sundays. We ask the user about this pattern after the study and they confirmed that they were only available during weekends.

\begin{figure*}[!tbp]
  \begin{center}
  \begin{minipage}[t]{.47\textwidth}
    \centering
    \includegraphics[width=1\textwidth]{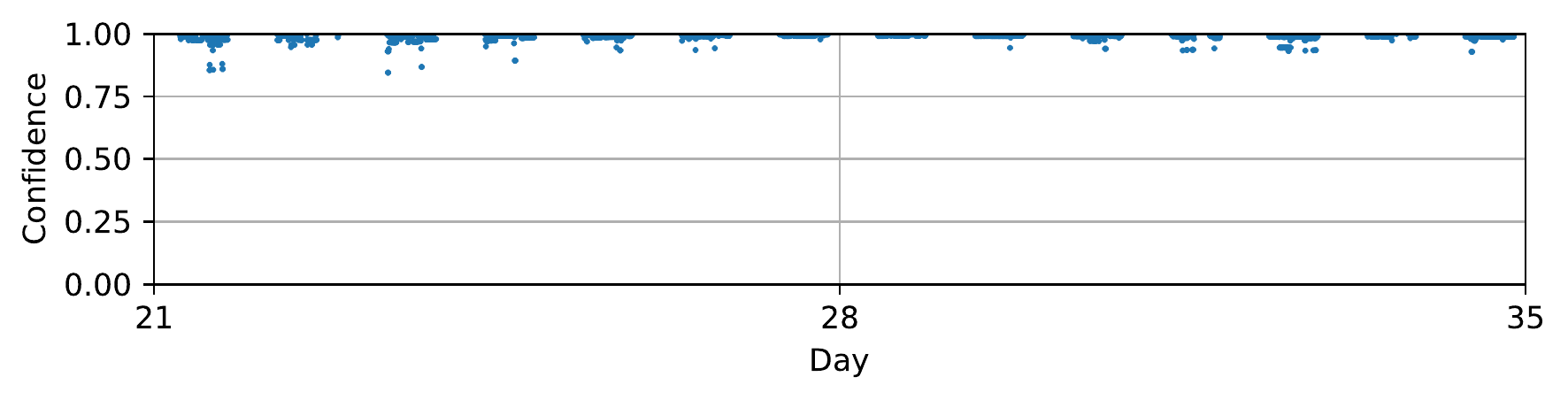}
    \small (a) Overall high confidence in RL (6 cases)
  \end{minipage}
  \begin{minipage}[t]{.47\textwidth}
    \centering
    \includegraphics[width=1\textwidth]{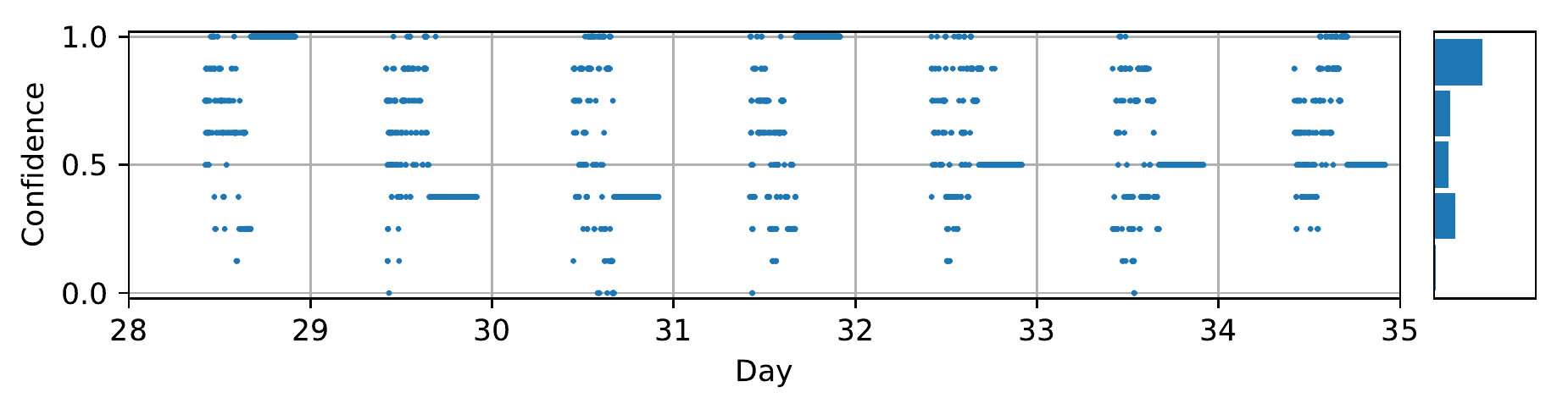}
    \small (e) Overall high confidence in SL (4 cases)
  \end{minipage}
  
  \begin{minipage}[b]{.47\textwidth}
    \centering
    \includegraphics[width=1\textwidth]{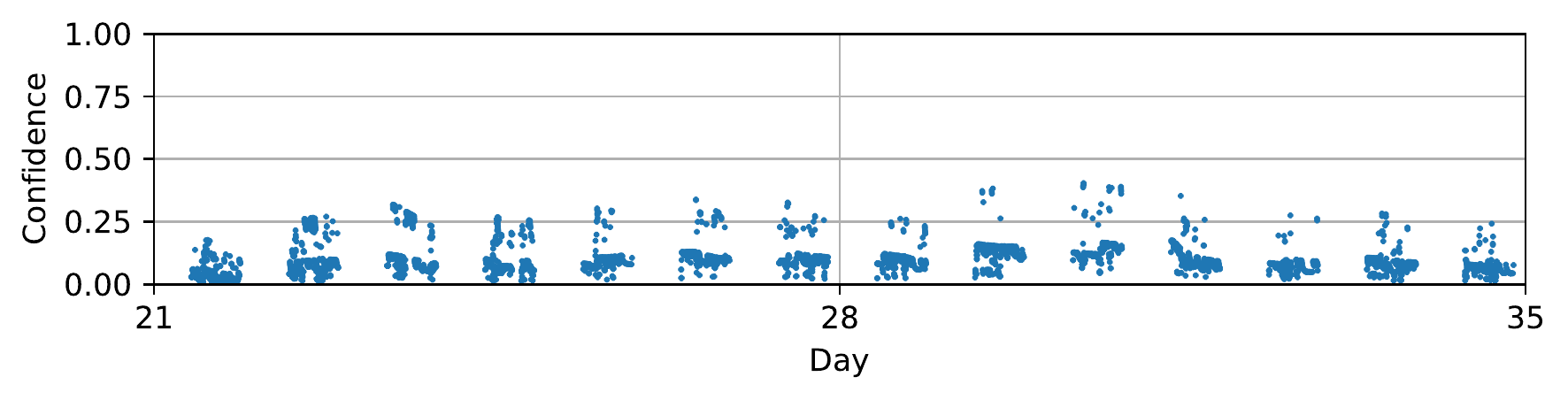}
    \small (b) Overall low confidence in RL (4 cases)
  \end{minipage}
  \begin{minipage}[b]{.47\textwidth}
    \centering
    \includegraphics[width=1\textwidth]{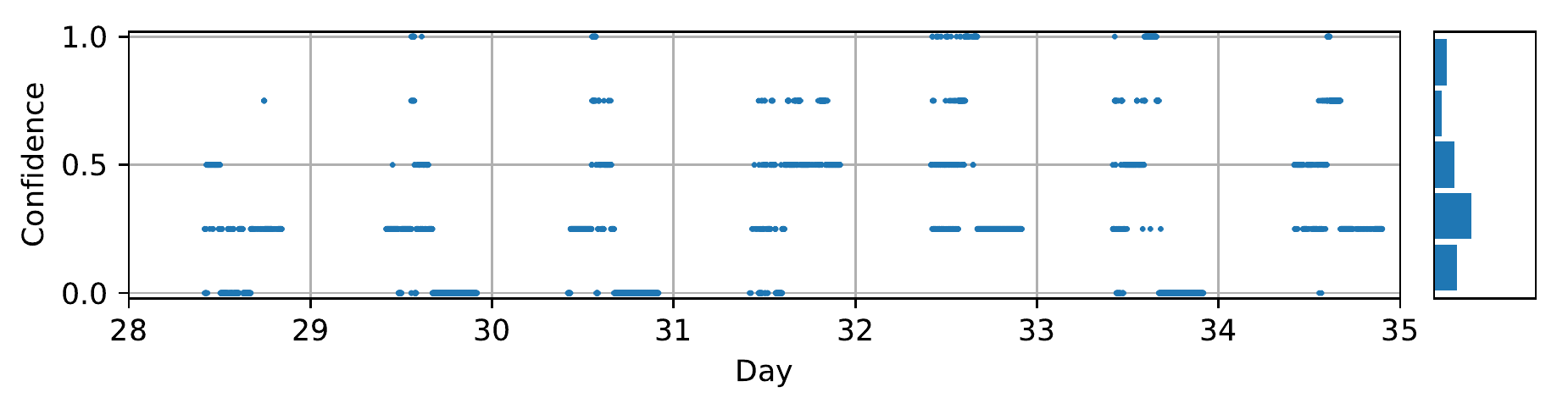}
    \small (f) Overall low confidence in SL (4 cases)
  \end{minipage}

  \begin{minipage}[b]{.47\textwidth}
    \centering
    \includegraphics[width=1\textwidth]{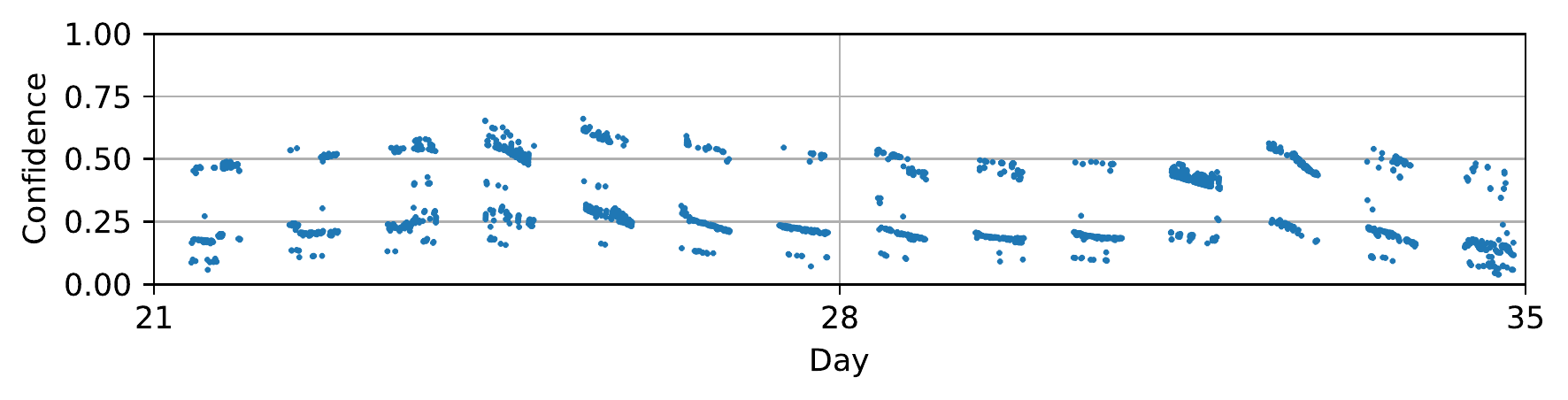}
    \small (c) Separated confidence in RL (3 cases)
  \end{minipage}
  \begin{minipage}[b]{.47\textwidth}
    \centering
    \includegraphics[width=1\textwidth]{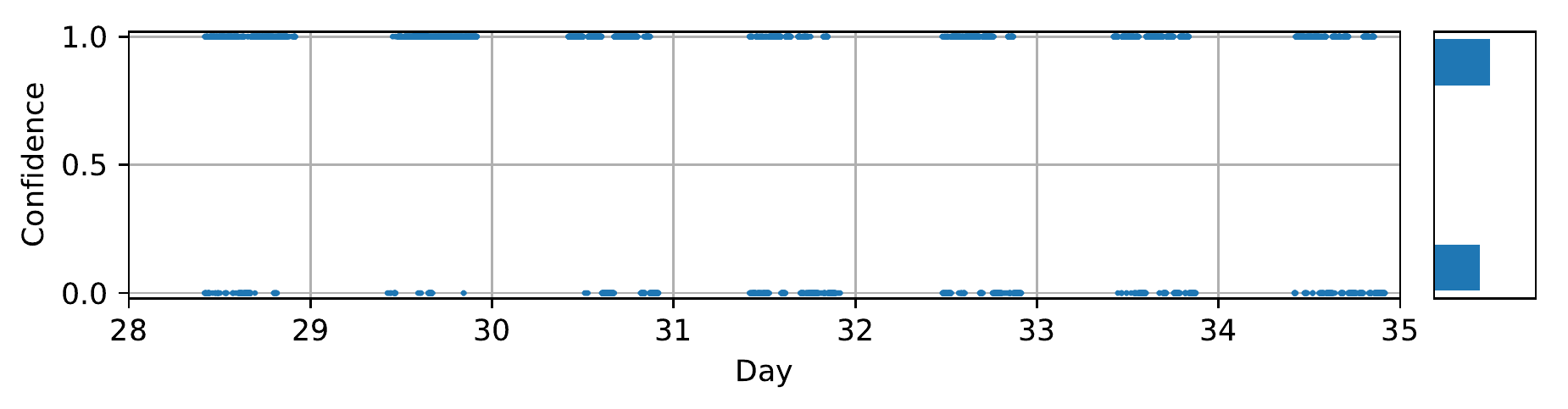}
    \small (g) Separated confidence in SL (3 case)
  \end{minipage}
  
  \begin{minipage}[b]{.47\textwidth}
    \centering
    \includegraphics[width=1\textwidth]{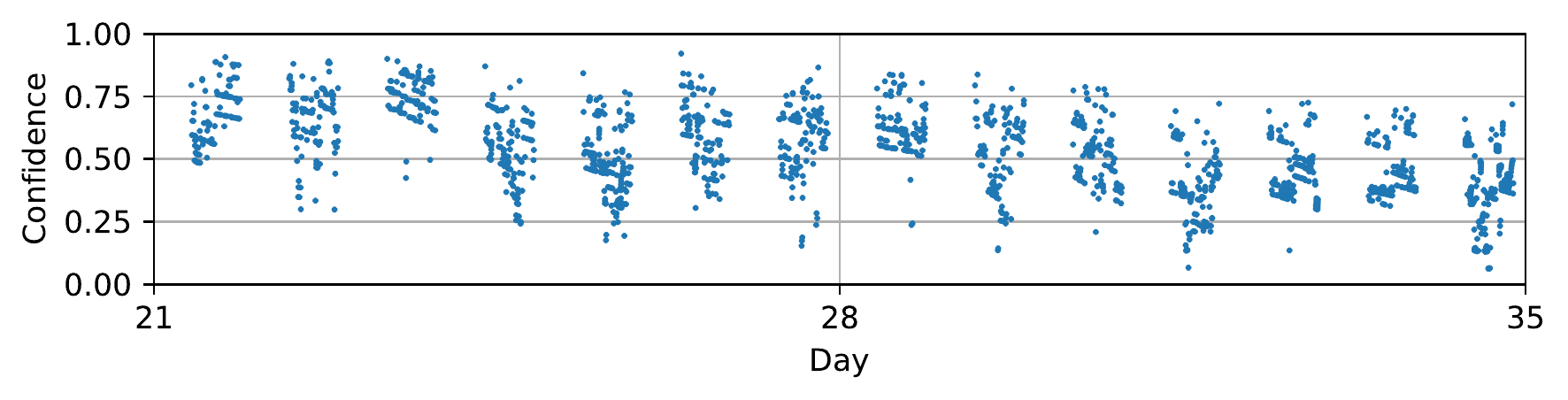}
    \small (d) Diffused confidence in RL (2 cases)
  \end{minipage}
  \begin{minipage}[b]{.47\textwidth}
    \centering
    \includegraphics[width=1\textwidth]{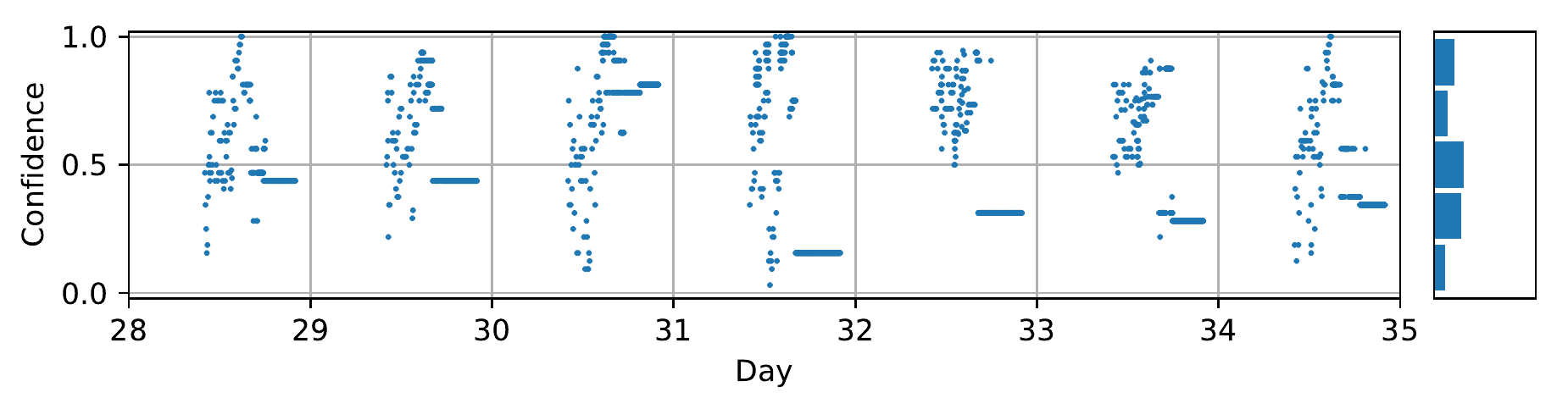}
    \small (h) Diffused confidence in SL (4 cases)
  \end{minipage}
  
  \caption{Examples of confidence change in RL (left column) and SL (right column). Four common patterns are found in both algorithms. For SL, we provide the confidence distribution on the side to assist visualization.}
  \label{fig:rl_sl_confidence_analysis}
  \vspace{-1em}
  \end{center}
\end{figure*}


\textbf{Users can be categorized into four coherent groups.} 
Notification response behavior varied widely between users. We analyzed the learned behaviors of both RL and SL agents for each user. For RL, we use the probability of sending a notification as given by the policy network. For SL, we use the Random Forest confidence, which is the number of votes it gets for sending a notification from the tree ensemble. 

Figure ~\ref{fig:rl_sl_confidence_analysis} depicts a typical user from the four groups we identified. 
The first group follows a high-confidence pattern because of the high answer rate, and the second follows a low-confidence pattern as users dismiss or ignore majority of the microtasks. The third pattern shows the points are vertically separated into two clouds. We observed that both RL and SL increase the confidence of sending a microtask when screen is turned on for these users, a manual rule used in prior work~\cite{vaish2014twitch}. The final pattern shows the confidence varies significantly within a day. Further analysis revealed several factors impact the decision. For example, we found that the agent becomes more confident when either the screen is on or ringtone mode is adjusted to normal in Figure~\ref{fig:rl_sl_confidence_analysis}d, and the confidence increases when screen is on or non-stationary motion is detected in Figure~\ref{fig:rl_sl_confidence_analysis}h. The combination of multiple variables cause different confidence levels. We checked the learned patterns with each of the users response patterns and they aligned well. 

We analyzed the rewards received in the testing phase per group. For the high-confidence group, RL receives $165.7\pm144.3$ (N=6) weekly rewards, and SL receives $-90.9\pm404.9$ (N=4). For low-confidence group, RL receives $-45.2\pm24.6$ (N=4) and SL receives $-99.1\pm75.6$ (N=4). While RL generally receives higher rewards in these groups, the results are not statistically significant due to high variance.

\subsection{System Performance}

\paragraph{Client App Battery Impact}
The battery consumption was measured on a new Pixel 2 phone running Android 8.0 with a sim card. We factory reset the phone to minimize measurement noise. We measured the battery consumption with and without our app installed separately. With a fully charged phone, our results show that the battery level drops to 84\% after 12 hours without our app installed, and to 77\% when our app is running in the background. Hence, our app increases battery use by 7\% during a day.

\paragraph{Server Request Handling}
Our server is hosted on a desktop with a 4-core Intel CPU @ 3.5 GHz and 32 GB DDR3 memory. We benchmarked the overhead of both algorithms when processing an interruptibility query. Supervised learning consumes 163 MB and takes $1.27 \pm 0.07$ seconds to complete a query, and RL consumes 243 MB and takes $2.28 \pm 0.16$ seconds. The major cause of the time overhead is for loading and initializing the agents in both algorithms. Our results suggest RL introduces higher overhead.

\subsection{Post-Study Survey}

We collected 26 effective post-study surveys, 14 from the RL group and 12 from the SL group. 
13 participants in RL group and 11 in SL observed a difference in notification patterns during the 5-week study. These users were subsequently asked to rate the change of the task schedule with a 5-point Likert scale where 1 is noticeably worse and 5 is noticeably better. The rating is $4.23 \pm 0.58$ and $3.45 \pm 1.44$ in RL and SL, respectively, implying that RL learns the opportune moment to engage as per user perception. Participants in RL express \emph{``started getting more/fewer notifications during specific times of the day''}. Participants mentioned receiving less undesired notifications during work (N=3)\footnote{We use N=? to denote number of people.}, studying (N=1), in the morning and evening (N=1), or receiving more notifications at  opportune times such as when they are ``sitting down'' (N=1). A few SL participants observed a polarizing change when transitioning to the testing phase: one participant received significantly more tasks, while two participants received significantly less amount of tasks (N=2). Two users indicated the app gave more notifications when they were studying (undesirable, N=2) while one user experienced less notifications when at work (N=1). 

RL and SL users expressed different concerns when asked why certain notifications were disruptive. RL participants indicated that the tasks were sent out far too often than they expected (N=6). Some reported that the app is unaware when they are engaged with other phone activities such as watching videos or playing games (N=5). On the other hand, the major concern of SL participants is that microtasks were delivered at inopportune moments in which they could not answer (e.g., driving, at work) (N=5). The frequency of notifications also heightens the disturbance (N=4).

In both algorithms, prompting at an inopportune moment is a major reason for dismissed notifications (N=8). Participants sometimes dismissed notifications when they found the microtasks to be too challenging (N=4). 16 users reported arithmetic questions to be more difficult than other questions (N=16). However, one user chose to randomly select answers instead of dismissing notifications (N=1).

\section{Discussion and Future Work}

The goal of an RL agent is to maximize the long term reward, and the reward function is designed to achieve the desired outcome. In \QuickQuestion, we investigate a simple objective which optimizes for the number of completed microtasks while minimizing the number of dismissed notifications to reduce disturbance. However, it is clear from our study that designing the reward function is non-trivial as the agent can have unintended behavior. In our case, the agent decided to send many notifications during the day as it was not penalized enough for ignored notifications.

We can improve the reward function as per application requirement. For example, the reward function can be augmented to discourage when a high-priority notification is missed. Also, our reward function can potentially incorporate the response rate and the response accuracy. In \QuickQuestion, we hand picked the reward ratio of answering and dismissing a notification to be 1 to 5, but a better reward mechanism can be explored based on behavioral models, or automatically optimized by Inverse Reinforcement Learning algorithms~\cite{banovic2017leveraging}. Designing a reward function that can generalize to all types of notifications is challenging, and a promising direction for future work. 

Although we keep the microtasks as homogeneous as possible in our study (i.e., length and task style), some questions do cause bias for certain users. For example, one user did not want to answer the diet questions because they made him feel self conscious. This bias, however, can be explicitly modeled in RL by augmenting the action space, i.e., the agent can decide which question to prompt based on user preference~\cite{greenewald2017action}. Another direction to be explored in the future is to consider a different workload based on the intensity of interruptibility~\cite{yuan2017busy}. For example, a system can prompt more than one microtasks in a row~\cite{cai2016chain} when a user is more available. 



\section{Conclusion}

We presented our system \QuickQuestion to understand the trade-off between supervised learning (SL) and reinforcement learning (RL) for identifying user interruptibility for microtasks. We conducted a 5-week user study with 30 participants to collect user interactions with notifications. Our results show that both SL and RL learn data driven patterns and identify interruptible moments manually discovered in prior work. RL and SL are commensurate in terms of overall performance as measured by rewards, and it is difficult to draw statistically significant conclusions due to high variance in user behavior. RL is more effective in reducing dismissed notifications as incentivized by its reward function. However, SL achieves a higher response rate, indicating that reward design plays an important role in guiding RL behavior. RL can smoothly adapt to changing user preferences and lower the burden for users to handle microtasks. Our user perception survey indicates that RL achieves better user experience and more accurately identifies interruptible moments. We open source our code and dataset to encourage future work in this area\footnote{\url{https://github.com/nesl/EngagementService}}.


\nocite{langley00}

\bibliography{ours}
\bibliographystyle{icml2020}





\clearpage
\section*{Appendix}
\appendix
\section{A2C Algorithm}
\label{appendix:algorithm}

The policy $\pi$ indicates which action $a_t$ should be performed given the current observation $o_t$. Let the discounted sum of future rewards $R_t = \sum_{k=0}^{\infty}\gamma^kr_{t+k}$ where $r_i$ is the reward received at the $i^{th}$ step and $\gamma$ is the discount factor between 0 to 1. The state-value function $V^\pi(o)$ is the expected value of discounted future rewards from a given observation $o$ following a policy $\pi$:
\begin{equation}
\label{eq:value_function_def}
V^\pi(o) = \mathbb{E} [ R_t  | o_t = o].
\end{equation}
Similarly, the action-value function is defined as the expected value of taking an action $a$ at an observation $o$ following a policy $\pi$:
\begin{equation}
\label{eq:q_function_def}
Q^\pi(o,a) =  \mathbb{E} [ R_t | o_t = o, a_t = a].
\end{equation}
Then, the advantage function which is defined as
\begin{equation}
\label{eq:advantage_function_def}
A^\pi(o,a) = Q^\pi(o,a)-V^\pi(o)
\end{equation}
indicates how advantageous it is to take an action at a given state compared to other actions. 

\QuickQuestion uses the Advantage Actor-Critic (A2C)~\cite{mnih2016asynchronous} as the RL algorithm. A2C uses two neural networks - an actor network and a critic network. The actor network generates actions by representing the policy $\pi(a|o, \theta)$ with parameters $\theta$, while the critic network learns the value function to assess the benefit of an action. 
The actor network outputs the probability distribution of actions (i.e., confidence), and critic network gives the feedback over the chosen action. 
In policy gradients methods, $\theta$ is updated in the direction of $\Delta_\theta \log \pi(a_t|o_t;\theta)R_t$ where $R_t$ is the accumulated reward after a policy run.  To reduce the variance of updates, an unbiased baseline is subtracted from the accumulated reward as $\Delta_\theta \log \pi(a_t|o_t;\theta)(R_t-b(t))$. In A2C, the baseline is the state-value function: $b(t)=V^\pi(o)$. Hence, the estimate of the value function as given by the critic network is used in computing the gradient.


\begin{equation}
\label{eq:value_function_recursive}
V^\pi(o_t) = \mathbb{E} [ r_t ] + \gamma \cdot V^\pi(o_{t+1})
\end{equation}

Equation (\ref{eq:value_function_recursive}) is a form of Bellman's equation and can be directly solved by dynamic programming. The drawback with this approach is that the state value $V(o)$ can be computed only when the observation $o$ is visited, hence limits the size of the state space.




In practice, we usually maintain one big network by merging the first few layers of both networks due to idential structure. For more details, please refers to the original paper~\cite{mnih2016asynchronous}.

 
\section{System Design and Implementation}
\label{appendix:system}


Figure~\ref{fig:system_summary}
shows our server-client architecture with: a phone client app and a web server. Our phone app senses user context data and sends it to the server every minute. Our server determines if the user is interruptible at the moment based on the current user context along with the past user daily routing and response history, and returns the binary decision back to the client app, indicating \emph{to prompt a microtask} (i.e., a short question) or \emph{to keep silent}. Once the app displays the microtask, the app tracks how the user responds to the question and sends the response by piggybacking on the next request.

\subsection{Client App}


The inclusion criteria of our study are that the participant is an active Android user and has the OS version 7 or higher installed. Android supports heads-up style notifications starting from Android 7, which is also the mainstream notification style in iOS. 

Our app is composed of three components: A \emph{sensing module} to monitor user context, a \emph{microtask pool} to store a list of short questions, and a \emph{user interface module}.

\paragraph{Sensing Module} 
It collects time of the day, day of the week, location, motion status, screen status, ringer mode and the elapsed time since last prompt (see  Table~\ref{tab1e:feature_list}). The challenge is to perform sensing continuously during the study period (i.e., 12 hours per day) while not drawing too much energy. 
To minimize battery impact, we use Android's adopt two strategies to save energy. First, we make the sensing process event-driven and use \texttt{AlarmManager} to schedule sensing tasks. This avoids unnecessary CPU idle time and keeps the CPU in sleep mode when possible. Second, we exploit 
use \texttt{ActivityRecognition} API for motion activities and the \texttt{Geofencing} API for location instead of collecting raw data. Both APIs are powered by Google Play Service which aims to optimize the sensing pipelines in the hardware- and operating-system level.

For motion, we collect activity labels (e.g., walking) instead of raw data. Similarly, for location, we ask participants to declare two geofences indicating their home and workplaces. If a user is neither at home nor at work, the location label is marked as \emph{others}. Since the app performs sensing continuously through 10am to 10pm, we exploit Google Play Service to collect motion activity and geofencing information which is optimized in the OS level to reduce the battery impact.

\paragraph{Microtask Pool}
Once the server receives the user context, the server informs the client app whether the user is available to complete a \emph{microtask}. A microtask can be anything that requires user input. In our system, we focus on 
We use ecological momentary assessment (EMA) questions, commonly used in human behavioral studies~\cite{stone1994ecological}. Inspired by~\citet{ponnada2017microinteraction} who argue that questions require short amount of time to complete increases the response rate, all the questions are designed in the micro-EMA style in multiple choice form and can be answered within a few seconds. 

The microtask pool stores the questions that the commercial apps may be interested. These questions can be partitioned into: (i) Self-monitoring questions that track user's mental and physical status such as stress and diet; (ii) Participatory sensing questions collect the environmental information, e.g., noise level at the current location; (iii) The crowdsourcing questions, e.g., image ground truth labeling. We have nine question types listed in Table~\ref{tab1e:question_list}. We developed several types of questions to simulate when different applications are interested in getting user responses. In these questions, the image question and arithmetic question have the correct answers that we can track the accuracy the responses.

\begin{figure}[t!]
  \centerline{\includegraphics[width=0.33\textwidth]{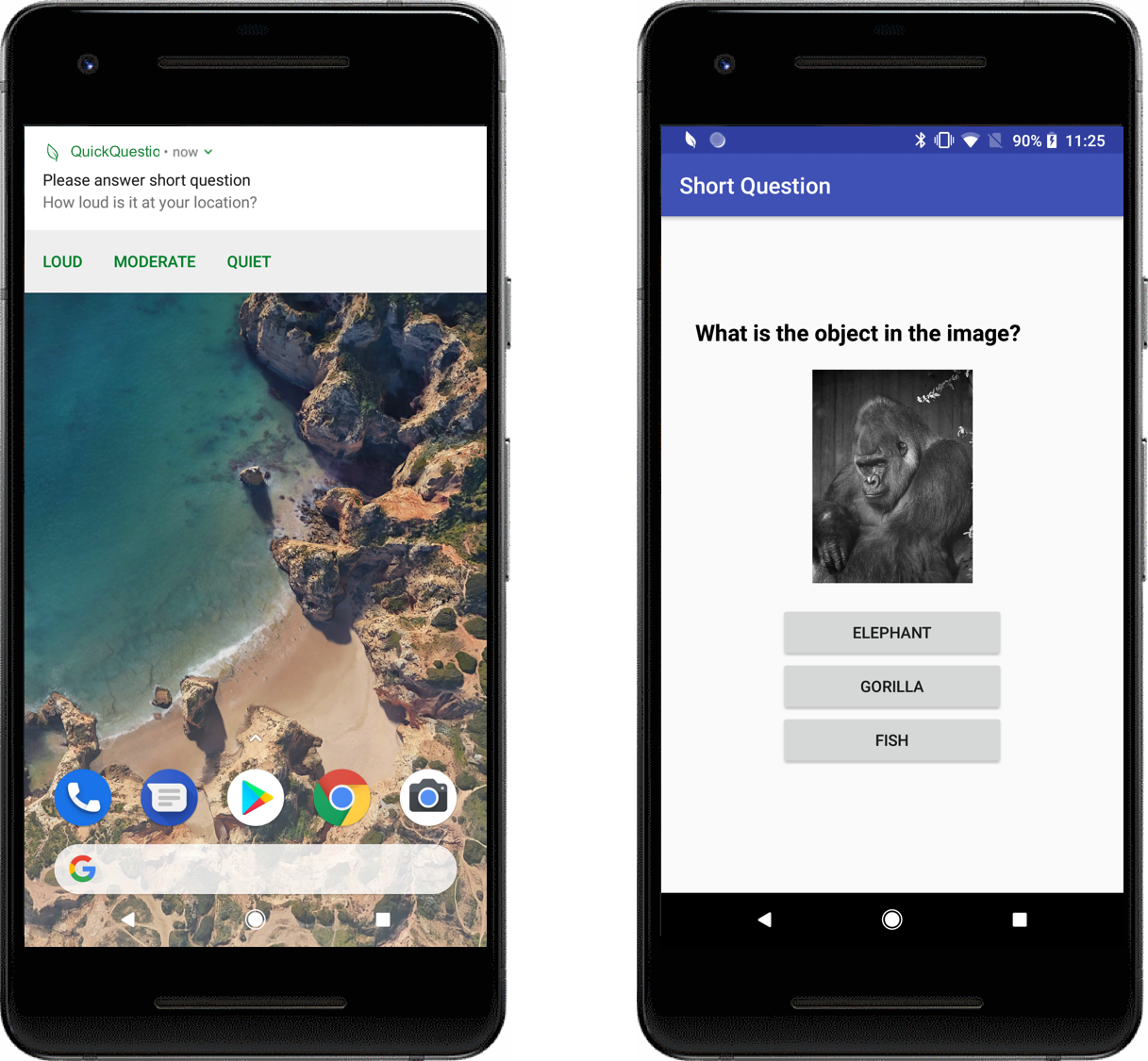}}
  \caption{The microtask answering interface in the client app, through a notification (left) or in the app (right).}
  \label{fig:screenshot_appendix}
\end{figure}

\paragraph{Notification interface}
We embed the questions into the push notification (Figure~\ref{fig:screenshot}). The notification is displayed \emph{heads-up} style with the possible options right below. Android supports heads-up style notifications starting from Android 7, which is also the mainstream notification style in iOS.
Alternatively, users can also answer the questions by manually launching our app and selecting a choice in the task view. A microtask times out after an hour, or when a new microtask is scheduled.

\subsection{Server}

Our server is composed of a standard \emph{web server}, a \emph{database system}, and several \emph{learning agents}. When the web server receives a request (i.e., the user reaction of the previous action and the current user context), it restores the learning agent by retrieving the \emph{learning policy} from the database. The agent updates the policy by considering the user reaction if necessary. The agent then makes an inference of user interruptibility and decides if it is appropriate to prompt a task at the moment based on the revised policy. The policy is dumped back to the database.
The server converts the user response and the user context from the request into a reward value and a state, and pass them to the RL worker. The RL worker processes the reward and decide which action to take based on the current state. The web server wraps the action as the response and sends it back to the client. The reason that we separate the learning agents from the web server is that the web server is stateless. However, reinforcement learning is a sequential process, hence, we have a standalone worker to preserve the sensing context.

\paragraph{HTTPS Server}
It serves as the frontend for our mobile client app to query user interruptibility through the RESTful API. We implement a dashboard to identify if the data is not collected as anticipated due to a connection loss from the user side. We use Django\footnote{Django Web Framework: https://www.djangoproject.com/} to develop our web application, which follows the standard model-view-controller (MVC) design pattern.

\paragraph{Database}
The database stores logs including the interruptibility request records, task response time and results, and the sensor data. These logs are for data analysis  and are not part of the interruptibility inference. The \emph{policy} of each learning agent is also kept in the database, which hosts a collection of learning workers. When a user registers herself to the system, our system spawns a learning worker. When the server receives a request from a client via HTTPS protocol, the server retrieves the state of the user and identifies which learning worker takes in charge of it. 

There are two types of learning workers: Supervised learning worker and reinforcement learning worker. Depending on which group the user is assigned to, the server then passes the user response for the previous action and the current user state to the corresponding learning worker.
\section{Hyperparameters}
\label{appendix:hyperparameters}

Below we provide the hyperparameters used for the experiment using both algorithms. Note one model is trained per participant in both algorithms.

\subsection{Reinforcement Learning (RL)}

No formal hyperparameter search was conducted and the hyperaparameters were generally set to default values found in the Intel Coach library~\cite{caspi_itai_2017_1134899}. 

\begin{table}[h]
	\centering
	
	\bgroup
	\def\arraystretch{1.1}
	\caption{Hyperparameters used for RL experiments. We used the Advantage Actor Critic algorithm, as implemented in Intel Coach repository~\cite{caspi_itai_2017_1134899}. We use the same network architecture for both actor and critic.}
	\label{table:rl-hyperparams}
	
	\begin{tabular}{|c|c|}
		\hline
		\textbf{Hyperparameter}  & \textbf{Value}  \\ \hline
		Discount factor         & 0.99            \\ \hline		
		SGD iterations & 5 \\ \hline
		Minibatch size & 64 \\ \hline
		Consecutive Playing Steps  & 512 \\ \hline		
		Consecutive Training Steps  & 1 \\ \hline		
		Learning rate & 0.0001 \\ \hline		
		Hidden layers         & [256] \\ \hline
        Use GAE         & True \\ \hline
        GAE Lambda      & 0.95 \\ \hline
        Clip Gradient   & 40 \\ \hline
        Normalize Observation & True \\ \hline
        Heatup Steps    & 0 \\ \hline
	\end{tabular}
	\egroup
\end{table}

\subsection{Supervised Learning (SL)}

A Random Forest Classifier is trained using the first 3 weeks of each participant's data. We then perform a grid search to pick the parameter set that yields the highest reward. Below we provide the possible parameter values that are considered. For other hyperparameters, we use the default value provided in Scikit-Learn library~\cite{scikit-learn}.

\begin{table}[h]
	\centering
	
	\bgroup
	\def\arraystretch{1.1}
	\caption{Hyperparameters used for SL experiments. We used the Random Forest algorithm provided in Scikit-Learn library~\cite{scikit-learn}}
	\label{table:sl-hyperparams}
	
	\begin{tabular}{|c|c|}
		\hline
		\textbf{Hyperparameter}  & \textbf{Possible Values}  \\ \hline
		n\_estimators         &  1, 2, 4, 8, 16, 32            \\ \hline		
		max\_features &   auto, sqrt, log2 \\ \hline
	\end{tabular}
	\egroup
\end{table}

\end{document}